\documentclass[fleqn,usenatbib]{mnras}

\usepackage{newtxtext,newtxmath}

\usepackage[T1]{fontenc}

\DeclareRobustCommand{\VAN}[3]{#2}
\let\VANthebibliography\thebibliography
\def\thebibliography{\DeclareRobustCommand{\VAN}[3]{##3}\VANthebibliography}

\usepackage{graphicx}	
\usepackage{amsmath}	
\usepackage{cleveref}
\usepackage{multirow}

\newcommand{\name}{\textsc{globalemu}}
\newcommand{\cmGEM}{\textsc{21cmGEM}}
\crefformat{figure}{Fig.~#2#1#3}
\crefformat{equation}{equation~(#2#1#3)}
\crefformat{section}{section~#2#1#3}
\crefformat{table}{Tab.~#2#1#3}
\crefformat{appendix}{appendix~#2#1#3}
\crefmultiformat{figure}{Figs.~#2#1#3}{~and~#2#1#3}%
    {,~#2#1#3}{,~#2#1#3}
\crefrangeformat{figure}{Figs.~(#3#1#4--#5#2#6)}

\title[\name: Emulating the Global 21-cm signal]{\name: A novel and robust approach for emulating the sky-averaged 21-cm signal from the cosmic dawn and epoch of reionisation}

\author[H. T. J. Bevins et al.]{
H. T. J. Bevins,$^{1}$\thanks{E-mail: htjb2@cam.ac.uk}
W. J. Handley,$^{1, 2}$
A. Fialkov,$^{2,3}$
E. de Lera Acedo$^{1, 2}$
and K. Javid$^{1, 2}$
\\
$^{1}$Astrophysics Group, Cavendish Laboratory, J. J. Thomson Avenue, Cambridge, CB3 0HE, UK\\
$^{2}$Kavli Institute for Cosmology, Madingley Road, Cambridge CB3 0HA, UK \\
$^{3}$Institute of Astronomy, University of Cambridge, Madingley Road, Cambridge CB3 0HA, UK
}

\date{Accepted XXX. Received YYY; in original form ZZZ}

\pubyear{2021}

\begin{document}
\label{firstpage}
\pagerange{\pageref{firstpage}--\pageref{lastpage}}
\maketitle

\begin{abstract}
Emulation of the Global (sky-averaged) 21-cm signal with neural networks has been shown to be an essential tool for physical signal modelling. In this paper we present \name, a Global 21-cm signal emulator that uses redshift as a character defining variable alongside a set of astrophysical parameters to estimate the signal brightness temperature. Combined with physically-motivated data pre-processing this makes for a reliable and fast emulator that is relatively insensitive to the network design. \name~can emulate a high resolution signal in $1.3$~ms in comparison to $133$~ms, a factor of $102$ improvement, when using the existing public state of the art \cmGEM. We illustrate, with the standard astrophysical models used to train \cmGEM, that \name~is almost twice as accurate and for a test set of $\approx1,700$ signals we achieve a mean $RMSE$ of $2.52$ mK across the band $z = 7 -28$ ($\approx$10 per cent the expected noise of the Radio Experiment for the Analysis of Cosmic Hydrogen (REACH)). The models are parameterised by the star formation efficiency, $f_*$, minimum virial circular velocity, $V_c$, X-ray efficiency, $f_X$, CMB optical depth, $\tau$, the slope and low energy cut off of the X-ray spectral energy density, $\alpha$ and $\nu_\mathrm{min}$, and the mean free path of ionizing photons, $R_\mathrm{mfp}$. \name~provides a flexible framework for easily emulating updated simulations of the Global signal and in addition the neutral fraction history. The emulator is pip installable and available at: \url{https://github.com/htjb/globalemu}. \name~will be used extensively by the REACH collaboration.
\end{abstract}

\begin{keywords}
dark ages -- reionization -- first stars -- early universe -- software: data analysis -- software: simulations
\end{keywords}



\section{Introduction}

The Global 21-cm signal from the Cosmic Dawn~(CD) and Epoch of Reionization~(EoR), if observed, will provide detailed information about the large scale properties of the early universe. The observable signal is the sky averaged 21-cm emission from the spin flip transition in neutral hydrogen at redshifts $z = 5-50$ and redshifted to frequencies of approximately $\nu = 50 - 200$ MHz.

An absorption signal was reported at $78$~MHz by the Experiment to Detect the Global Epoch of Reionization Signature~(EDGES) collaboration in 2018 \citep{Bowman2018}. However, the reported signal is significantly larger in amplitude than that predicted by standard $\Lambda$CDM cosmology \citep{Reis2021} and there are concerns about the data analysis used~\citep{Hills2018, Singh2019, Sims2020, Bevins2021}. Efforts are underway to make further observations of the signature with a variety of different radio telescopes including SARAS~\citep[Shaped Antenna measurement of the background RAdio Spectrum,][]{Singh2018}, REACH~\citep[Radio Experiment for the Analysis of Cosmic Hydrogen,][]{Acedo2019}, PRIZM~\citep[Probing Radio Intensity at High-Z from Marion,][]{Philip2019}, LEDA \citep[Large-aperture Experiment to Detect the Dark Ages,][]{Price2018}, DAPPER (Dark Ages Polarimeter PathfindER, \url{https://www.colorado.edu/project/dark-ages-polarimeter-pathfinder/}) and MIST (Mapper of the IGM Spin Temperature, \url{http://www.physics.mcgill.ca/mist/}) among others.

The intensity of the signal is measured against the radio background, typically assumed to be equal to the CMB temperature, and characterised by an absorption trough and an emission at late redshifts. The relative magnitude of the signal features is determined by various astrophysical processes including; the Wouthuysen-Field effect~\citep{Wouthuysen1952, Field1959}, Lyman-$\alpha$ heating and CMB heating \citep{Chuzhoy2007, Venumadhav2018, Villanueva2020, Mittal2020, Reis2021}, X-ray heating and ionization of the hydrogen gas by UV emission \citep{Madau1997}. A detailed discussion of the physics describing the Global 21-cm signal can be found in \cite{Furlanetto2006}, \cite{Pritchard2012}, \cite{Barkana2016} and \cite{Mesinger2019}. The physical processes themselves and hence the Global signal can be characterised by a set of astrophysical parameters (see \cref{training data} and \cite{Cohen2020}): the star formation efficiency, $f_*$, the minimal virial circular velocity, $V_c$, the X-ray efficiency, $f_X$, the CMB optical depth, $\tau$, the slope of the X-ray spectral energy density~(SED), $\alpha$, the low energy cut off of the X-ray SED, $\nu_\mathrm{min}$, and the mean free path of ionizing photons, $R_\mathrm{mfp}$.

Hybrid approaches are used to calculate realizations of the 21-cm signal, over a large cosmological volume and redshift range, which then can be averaged at every redshift separately to give the Global signal \citep[e.g.][]{Mesinger2011, Visbal2012,Fialkov2014, Cohen2017, Reis2021}. Each simulation takes several hours to perform on a desktop \citep{Monsalve2019} and though this is much faster than hydrodynamical simulations this time is too long to allow us to constrain astrophysical parameters using data. Therefore the desire to emulate the Global signal with neural networks, trained on the results of the large scale simulations, has arisen. The neural networks can produce a realisation of the Global signal in a fraction of a second by interpolating between the simulated cosmological and astrophysical models. This means they can be used to physically model the signal in, for example, Bayesian nested sampling loops\footnote{Note that here we are not referring to a Bayesian Neural Network \citep[see][]{Javid2020} but rather parameter optimisation algorithms such as \textsc{polychord} \citep[see][]{Handley2015a, Handley2015b}.}, as in the REACH data analysis pipeline~\citep{Anstey2020}, in which millions of calculations need to be made to infer cosmological parameters~\citep[][Sims et al. 2021 (in prep.)]{Liu2020, List2020, Chatterjee2021}. 

A number of papers have considered emulation of the 21-cm power spectrum using convolutional neural networks and other techniques \citep[e.g. ][]{Schmit2018, Jennings2018, Mondal2020}. At the time of writing \cmGEM~ is the only publicly available emulator used to accurately, with a maximum normalised RMSE of $10.55$ per cent, and quickly (see \cref{timing}) emulate the Global 21-cm signal \citep{Cohen2020} \footnote{During review of this work a pre-print describing the global signal emulator \textsc{21cmVAE} has been published \citep{21cmVAE} in which the comparative performance of \name~is discussed.}. It has previously been used to provide constraints on the parameter space of the 21-cm signal using EDGES high-band data \citep{Monsalve2019}. The emulator uses Principle Component Analysis~\citep[PCA,][]{Pearson1901}, the seven astrophysical parameters detailed above and in \cref{training data}, additional information about the mean collapsed fraction of halos as a function of redshift, $f_\mathrm{coll}(z)$, the fraction of X-ray energy above 1 keV, $f_{\mathrm{XR} > 1 \mathrm{keV}}$, and 2 keV, $f_{\mathrm{XR} > 2 \mathrm{keV}}$, and relies on a division of the signal into 2 or 3 distinct segments defined by the turning points. It involves the application of a decision tree for classification and several regression neural networks estimating PCA components, the frequencies and temperatures of turning points as well as additional parameters such as the frequency at which the neutral fraction equals 0.16, $\nu(x_{HI} = 0.16)$.

\begin{figure*}
    \centering
    \includegraphics{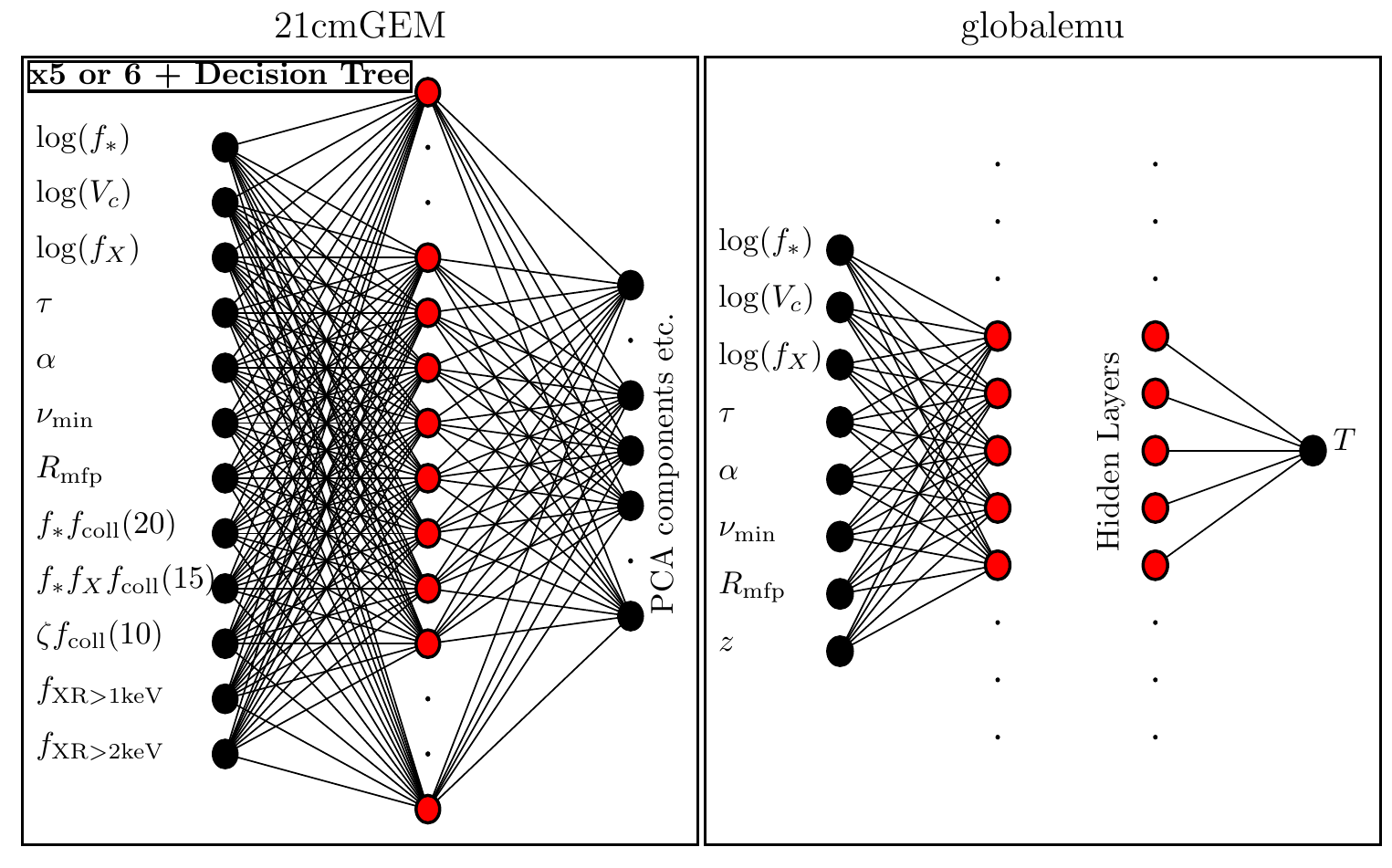}
    \caption{\textbf{Left Panel:} There currently exists only one other global signal emulator, \cmGEM, and we provide here an illustration of the regression neural networks used there. Note that \cmGEM~uses either 5 or 6 of these networks and a decision tree when making predictions.  For one of the regression networks used the only input parameters are the seven astrophysical parameters detailed in \cref{training data}~(excluding redshift). However, for the others there is an additional five derived parameters (see text and \protect\cite{Cohen2020}) and we illustrate the full set of 12 inputs. The number of output nodes depends on the specific application of the network and they can correspond to either PCA components (4 nodes), additional parameters such as $\nu(x_{HI} = 0.16)$ (1 node) and the frequencies and brightness temperatures of turning points in the signal (7 or 5 nodes). The hidden layer in all of the \cmGEM~regression networks has 40 nodes. For a full illustration of the \cmGEM~algorithm and detailed description see Fig. 11 and section 4.2 in \protect\cite{Cohen2020}. \textbf{Right Panel:} An illustration of the \name~neural network. Note the use of only one network to emulate the global signal in comparison to the 5 or 6 used for \cmGEM. Here the input layer has eight nodes (seven astrophysical parameters plus redshift) and the output layer is a single node returning the brightness temperature corresponding to the input redshift. We show a sizable hidden layer structure here with the red nodes and `...' however we note that the reduced number of inputs and outputs implies that a small architecture will be sufficient to achieve a high level of accuracy in the emulation~(see \cref{structure}).}
    \label{fig:network_types}
\end{figure*}

In this paper we present \name~which uses a novel and robust approach with a single small scale neural network to emulate the Global 21-cm signal given a comprehensive set of astrophysical parameters and redshift range. Where previously \cmGEM~was designed to take in astrophysical parameters and return a low dimensional representation of the Global signal as a function of redshift, \name~takes in the same astrophysical parameters and redshift and returns the signal temperature at the corresponding redshift~(see \cref{fig:network_types}). This greatly simplifies the complexity of the relationship being learnt by the neural network. It means we can achieve accurate results, with the smoothness of the signal imposed by the interpolation of the neural network between signals in the training data set, using a small network and without the need for a compressed representation of the signals like PCA where there is a potential loss of information. Additionally, \name~relies on a physically motivated pre-processing of the training data and can emulate a high resolution, $\delta z = 0.1$ over the range $z = 5 - 50$, Global 21-cm signal in $\approx1.3$ ms. \name~will be used extensively by the REACH collaboration and has been designed to have an average accuracy less than or equal to 10 per cent of the expected noise in the REACH system, estimated at ~25 mK (REACH Collaboration 2021 (in prep.)).

\name~is written in Python using \texttt{tensorflow} and the \texttt{KERAS} backend, it is pip installable via \texttt{pip install globalemu} and available at \url{https://github.com/htjb/globalemu}. It is flexible enough to be retrained on any set of Global 21-cm signal models whilst maintaing the novel design and physically motivated pre-processing. We provide a demonstration of its accuracy and efficiency in this paper using the same data used to train \cmGEM~and the corresponding trained models are publicly available on GitHub. 
We use GitHub actions to perform continuous integration.

In \cref{parameterisation} we describe the novel approach used to parameterise \name. \Cref{training data} describes the training and test data used to illustrate the capabilities of \name~in this paper and the astrophysical parameters in the simulations of the Global signal. We then describe the predominantly physically motivated pre-processing of the inputs and outputs of the neural network in \cref{preprocessing}. A discussion of the neural network structure follows in \cref{structure} and the quality of emulation is assessed in \cref{results}. We conclude in \cref{conclusions}.

\section{Parameterising the Problem}
\label{parameterisation}

There are several approaches that can be used to emulate the Global 21-cm signal with a neural network. The ultimate goal of the emulator is to take in a set of astrophysical parameters and return an estimate of the signal brightness temperature as a function of redshift, $\delta T(z)$, where the relationship has been learned from detailed numerical simulations. This can be done directly with a neural network that returns a value of $\delta T$ for each redshift data point it has be trained on. However, assuming the network is trained on high resolution signals this would result in a large number of outputs, making it hard to train, and would be limited in predictive power to specific values of redshift. The process can also be achieved by estimating, via a neural network, coefficients of a compressed representation of the signal space. For example, using PCA as with \cmGEM~\citep{Cohen2020} or learning coefficients of basis functions for polynomials or wavelets that when combined return the Global signal. However, whilst this approach reduces the number of outputs compared to a direct emulation, if incorrectly designed this can result in information loss and is equally limited in predictive power.

We take the novel approach of using redshift as an input to the network alongside the astrophysical parameters and returning from the network a single temperature corresponding to the given redshift. This is beneficial for two reasons; the small number of inputs and outputs means that the network can retain a simple structure and second the network will be able to interpolate between the values of redshift that it has been trained on. The smooth structure of the output from the network is guaranteed by the smooth interpolation performed by the neural network and by the smooth structure of the signals it is learning. Vectorised calls to the network are used to emulate the temperature as a function of redshift.

In \name~we also provide the ability to emulate the evolution of the neutral fraction, $x_{HI}$, of hydrogen as a function of redshift. We use an identical framework as when emulating the Global signal to do this with a set of astrophysical parameters and a redshift as inputs to a neural network and a corresponding value of $x_{HI}$ as an output.

We have built a network that can emulate the Global 21-cm signal to a high degree of accuracy without the need for L1 and L2 regularisation, dropout \citep{Srivastava2014}, batch normalisation \citep{Ioffe2015} or other similar concepts. We have achieved this by focusing on the pre-processing of the networks inputs and outputs in the desire to set our problem up in a way that is simple to solve with a basic neural network of a `reasonable' size.

\section{The Training and Test Data}
\label{training data}

\begin{figure}
    \centering
    \includegraphics{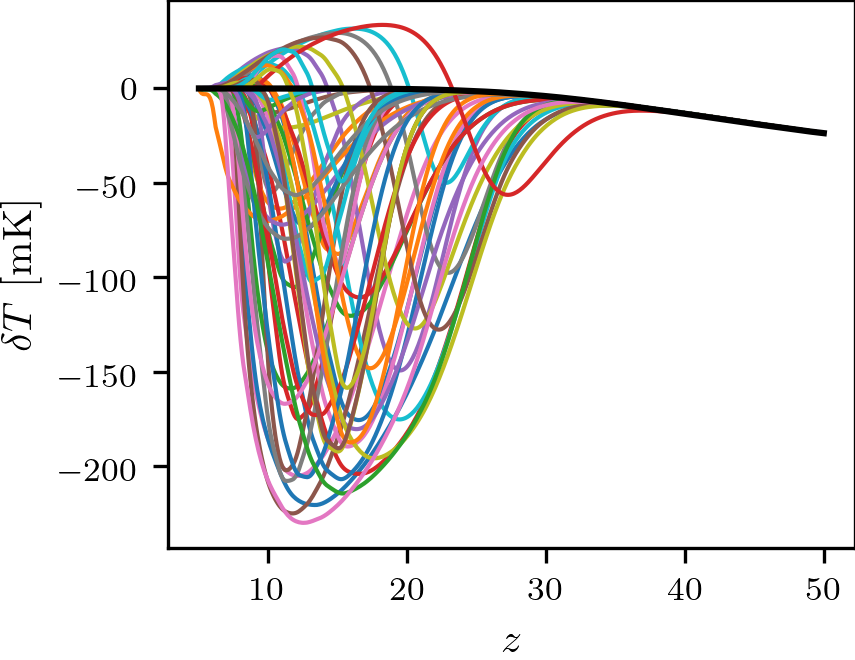}
    \caption{A subsample of 50 Global 21-cm signals from the \cmGEM~training set used here to demonstrate the efficiency of \name. The signals show the expected variety of structure with deep and shallow absorption troughs caused by Lyman-$\alpha$ coupling and terminated by X-ray heating. We also see emission against the CMB background at low redshifts in some of the models where there has been sufficient heating. Also shown in black is the Astrophysics Free Baseline~(AFB) (\cref{sec:afb}) which we model and remove from the training signals before we pass them through the neural network. Subtraction of the AFB prevents our network from attempting to learn a steadily decreasing temperature at high redshifts prior to star formation.}
    \label{fig:example_signals}
\end{figure}

In this paper we use the same model signals and corresponding astrophysical parameters used to train and test \cmGEM~\citep[available at \url{https://doi.org/10.5281/zenodo.4541500},][]{Cohen2021Data}. Examples of the Global signals from the training set are shown in \cref{fig:example_signals}. In total the data set contains 27,292 training models and 2,174 test models with each model being dependent on 7 astrophysical parameters. Each Global signal in the data set has 451 redshift data points and this means that each signal corresponds to 451 training points with the same astrophysical parameters and different redshits. Therefore, for \name~the 27,292 training models become 12,308,692 training data points. However, we continue throughout this paper to talk generally of training signals rather than data points because the emulator will be used to determine the signal structure over a redshift range~(see \cref{results} for more details).

The data pre-processing is explained in detail in \cref{preprocessing}. Explicitly, the inputs \citep{Cohen2017, Monsalve2019, Cohen2020} to the neural network when training on the \cmGEM~data are as follows (ranges are based on those in the \cmGEM~training data set, see section 2.5 of \cite{Cohen2020});
\begin{itemize}
    \item $f_*$: The star formation efficiency takes values in the range $0.0001 - 0.5$ and characterises the amount of gas converted to stars in the dark matter halos. A low star formation rate results in a low Lyman-$\alpha$ flux and late onset of X-ray heating leading to a shallower absorption trough and weak or non-existent emission above 0 mK in the signal.
    \item $V_c$: The minimal virial circular velocity has a value in the range $4.2 - 100$~km/s and is proportional to the cube root of the minimum threshold mass for star formation. A low value of $V_c$ corresponds to a small minimum mass threshold which in turn leads to an earlier onset of Lyman-$\alpha$ coupling, responsible for the absorption feature in the Global 21-cm signal, and shifts the minimum of the signal to higher redshifts.
    \item $f_X$: The X-ray efficiency of sources has a range between $0-1000$ and a high value corresponds to a high total X-ray luminosity. This leads to an earlier onset of X-ray heating which also shifts the minimum of the signal to higher redshifts, contributes to a shallower absorption and results in a significant emission feature during re-ionisation at low redshifts.
    \item $\tau$: The CMB optical depth in the \cmGEM~data sets takes a value in the range $0.04 - 0.2$ and a higher value of $\tau$ corresponds to a higher value of the ionizing efficiency of sources, $\zeta$. For high $\tau$ we would see an earlier reionisation of the hydrogen gas. We note that $\tau$ is given as $0.054 \pm 0.007$ by \cite{Planck2018} and that this falls at the lower end of the range in our training and testing data set. More recent parameter studies have explored lower values of $\tau$ in greater detail \citep{Reis2021}. However, the \cmGEM~data is sufficient to demonstrate the abilities of \name.
    \item $\alpha$: The power defining the slope of the X-ray SED with a range given by $1-1.5$. The Global 21-cm signal is expected to have a very weak dependence on $\alpha$ with the largest effect happening at low redshifts.
    \item $\nu_\mathrm{min}$: The low energy cut off of the X-ray SED has a range of $0.1-3$~keV. Low values of $\nu_\mathrm{min}$ correspond to a soft X-ray SED, efficient X-ray heating and a weak absorption feature in the 21-cm signal.
    \item $R_\mathrm{mfp}$: The mean free path of ionizing photons, with a range $10-50$~Mpc. $R_\mathrm{mfp}$ is expected to have a very weak effect and only at low redshifts \citep[see e.g. ][]{Monsalve2019}. A low $R_\mathrm{mfp}$ corresponds to a slower ionization of the neutral hydrogen gas.
    \item $z$: The redshift of the 21-cm brightness temperature is a measure of time and provides details about when each feature of the signal occurred. For example the brightness temperature is expected to reach 0~mK, corresponding to the end of the EoR, at low redshifts or more recent times. It is interchangeable with frequency given that the rest frequency, $\nu_r$, of the 21-cm line is 1420~MHz
    \begin{equation}
        z + 1 = \frac{\nu_r}{\nu}.
    \end{equation}
\end{itemize}

To ensure that we make a fair comparison of our results with those found when using \cmGEM~we make the same physically motivated cuts to the test data as are detailed in section 2.4 of \cite{Cohen2020}. This equates to limits on the ionizing efficiency of sources, $\zeta < \zeta_\mathrm{max} = 40,000 f_*$ and on the neutral fraction history at $z = 5.9$, $x_{HI}(z = 5.9) < 0.16$. Respectively the limits are motivated by stellar models \citep{Bromm2001} and quasar absorption troughs \citep{McGreer2014}. We also note that some of the parameters in the testing data have different ranges and the ranges are as follows; $f_*$: $0.0001 - 0.5$, $V_c$: $4.2 - 76.5$ km/s, $f_X$: $0-10$, $\tau$: $0.055 - 0.1$, $\alpha$: $1 - 1.5$, $\nu_\mathrm{min}$: $0.1 - 3$ keV and $R_\mathrm{mfp}$: $10 - 50$ Mpc. In total the final testing data set is comprised of 1703 models.

\name~includes a simple to use \textsc{python} graphical user interface~(GUI)\footnote{After installation via \textsc{pip} or from source the GUI can be called from the terminal using the command \textit{globalemu}. See the documentation at \url{https://globalemu.readthedocs.io/} for more details.} in which the variation of the signal with each of the astrophysical parameters listed above can be explored in more detail.
We note that the GUI is a feature made possible by the speed of emulation when using \name~(see \cref{results}).
There is an equivalent GUI for the neutral fraction history emulation.

As previously stated, \name~is not limited to emulating signals modelled with the above astrophysical parameters. It is flexible enough that more complicated astrophysical relationships can also be emulated. For example one explanation for the unexpected depth of the EDGES absorption trough is the presence of a higher than expected radio background which can be characterised with a quantity $f_\mathrm{radio}$ determining the normalisation of the radio emissivity~\citep[assuming the source of the excess radio background is stellar,][]{Reis2020}. \name~is in principle capable of being trained on models that consider $f_\mathrm{radio}$ in addition to the above 7 astrophysical parameters and redshift as inputs since it assumes nothing about the astrophysical parameters themselves. Equally \name~could be trained on less complex models.

\begin{figure}
    \centering
    \includegraphics{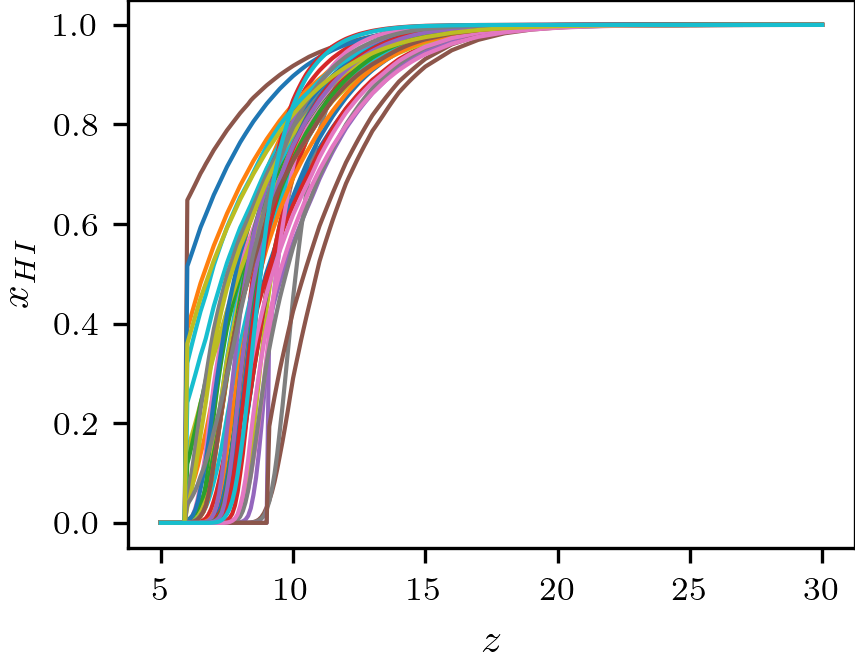}
    \caption{A subsample of 50 neutral fraction histories from the training set used in this paper. At high redshift the hydrogen in the universe is predominantly neutral and consequently $x_{HI}=1$. As the gas is ionised by UV emission from the first stars that form the neutral fraction decreases until $x_{HI}=0$ at the end of the EoR. }
    \label{fig:xhi_example_signals}
\end{figure}

For the neutral fraction, $x_{HI}$, we use a set of models produced as a by-product of the detailed \cmGEM~Global signal simulations. The data set is smaller with 10,047 training models and 791 test models however the relationship between the astrophysical parameters and $x_{HI}$ is expected to be (and shown to be, \cref{results}) simpler. We note that for $z \gtrsim 30$ the neutral fraction is expected to always be 1 and so we only emulate the neutral fraction over the range $z=5-30$. The models have not been released publicly but the parameter ranges are the same for this data set as detailed above. A subsample of the training models is shown in \cref{fig:xhi_example_signals}.

We note that a non-uniform coverage of the parameter space in the training data set, as with the \cmGEM~Global signal and neutral fraction data, may introduce bias in the neural network. The network will tend to learn regions of the astrophysical parameter space where the sampling is heavier better than others. For the purposes of illustrating the accuracy of the emulation in this paper this is not an issue. However, it can become an issue when using an emulator to physically model a signal in a data set where parameter estimation may be biased towards a false set of parameters. Training a network on a more uniform data set can alleviate this issue and we leave further exploration of this for future work.

\section{Data Pre-processing}
\label{preprocessing}

\begin{figure}
    \centering
    \includegraphics{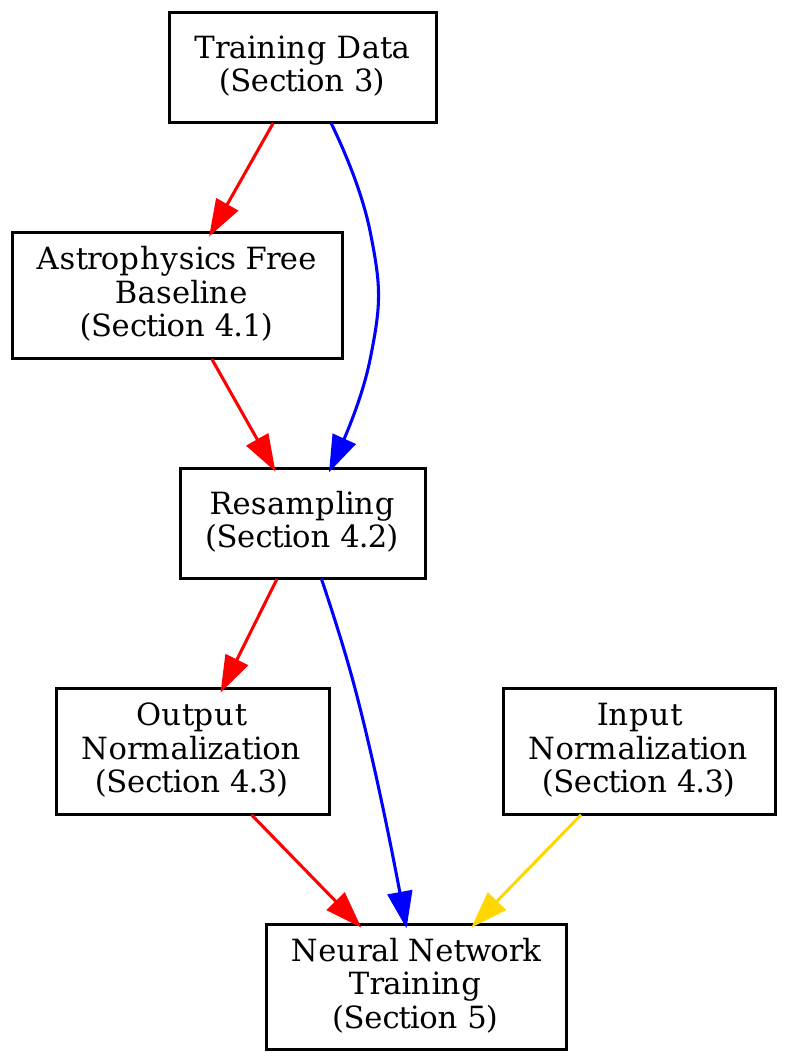}
    \caption{The pre-processing applied to the training data in \name. Each box is outlined in more detail in the corresponding sections. The red path is the pre-processing steps used for the Global 21-cm Signals, the blue path for the neutral fraction histories and the gold path are steps that occur when training both neural networks.}
    \label{fig:flow}
\end{figure}

The details in the following discussion\footnote{Specifically, the discussion details the steps used in this paper when training with the \textsc{21cmGEM} data. However, the various pre-processing steps outlined can be switched on and off by a user when training an emulator with \textsc{globalemu}.} outline the pre-processing for the network predicting the Global 21-cm signal. In \cref{neutral frac pp} we briefly discuss the pre-processing for the neutral fraction networks which is a largely similar process. The pre-processing is summarised as a flow chart in \cref{fig:flow}.

\subsection{Astrophysics Free Baseline Subtraction}
\label{sec:afb}

In the region where the structure of the Global 21-cm signal is expected to be dominated by collisional coupling it is independent on the 7 astrophysical parameters used here as inputs to the emulator. This means that, in the corresponding redshift range, each of the signals in our training and testing data sets have the same brightness temperatures. To prevent our network unnecessarily attempting to learn a non-trivial structure in this region we can treat it as an Astrophysics Free Baseline~(AFB), model and remove it from the signals before they are passed to the network for training. By doing this our network will learn a simpler relationship at high redshift between the parameters and $\delta T(z)$ than the existing steadily decreasing trend (see \cref{fig:example_signals}).

In \cref{appendixA} we give an approximate calculation of the AFB for the simulated signals that comprise the training data sets. The calculation is approximate because it follows the mean evolution of the signal and in contrast the simulations are produced over large-scale cosmological volumes evolved over cosmological time then averaged. We therefore normalise our result to the temperature of the signals in the training set at $z = 50$ and find that this is sufficient to represent the astrophysics free component of the models.

As stated, the AFB is then subtracted from the models before training the network and added back in after making predictions.

\cmGEM~uses five extra parameters in addition to the seven astrophysical parameters used here. In principle these parameters could be passed to the neural network during training due to the flexible nature of \name. However, three of these additional parameters rely on the fraction of mass contained in halos above the minimum cooling threshold, $f_\mathrm{coll}(z)$, and help the network learn the signal structure at high redshift where collisional coupling (cosmology) dominates. Here we do not consider these parameters, which are derived from the seven astrophysical parameters described above, as we are instead removing the AFB. The final two parameters, for reference, are the fractions of X-ray energy above 1 keV and 2 keV. These parameters are added to further characterise the X-ray SED but we find with \name~that we do not need to consider them to achieve accurate results.

\subsection{Resampling of signals}

The turning points, and gradients between them, of the Global 21-cm signal encode all of the information about the efficiency of Lyman-$\alpha$ coupling, X-ray heating and reionization. They are therefore highly dependent on the relevant astrophysical parameters and in the region where the features typically occur variation in the signals is significant. The original signal models are sampled uniformly in redshift and there is not particular physical motivation for this. However, to improve the quality of modelling we resample the signals at a higher rate across the redshift ranges that typically correspond to the locations of the turning points and at a lower rate where the signal structure deviates less from the `average' signal (e.g. above $z\approx30$ where the signal is free of astrophysics).

To do this we look at the variation in the signal amplitudes, after subtracting the AFB, across the training data set
\begin{equation}
    \Delta (\delta T(z)) = \delta T_\mathrm{max}(z) - \delta T_\mathrm{min}(z),
    \label{eq:resamp-dif}
\end{equation}
and we treat this as a probability distribution
\begin{equation}
    P(\Delta (\delta T(z))) = \frac{\Delta (\delta T(z))}{\sum_z \Delta (\delta T(z))}.
    \label{eq:resamp-prob}
\end{equation}

\begin{figure}
    \centering
    \includegraphics{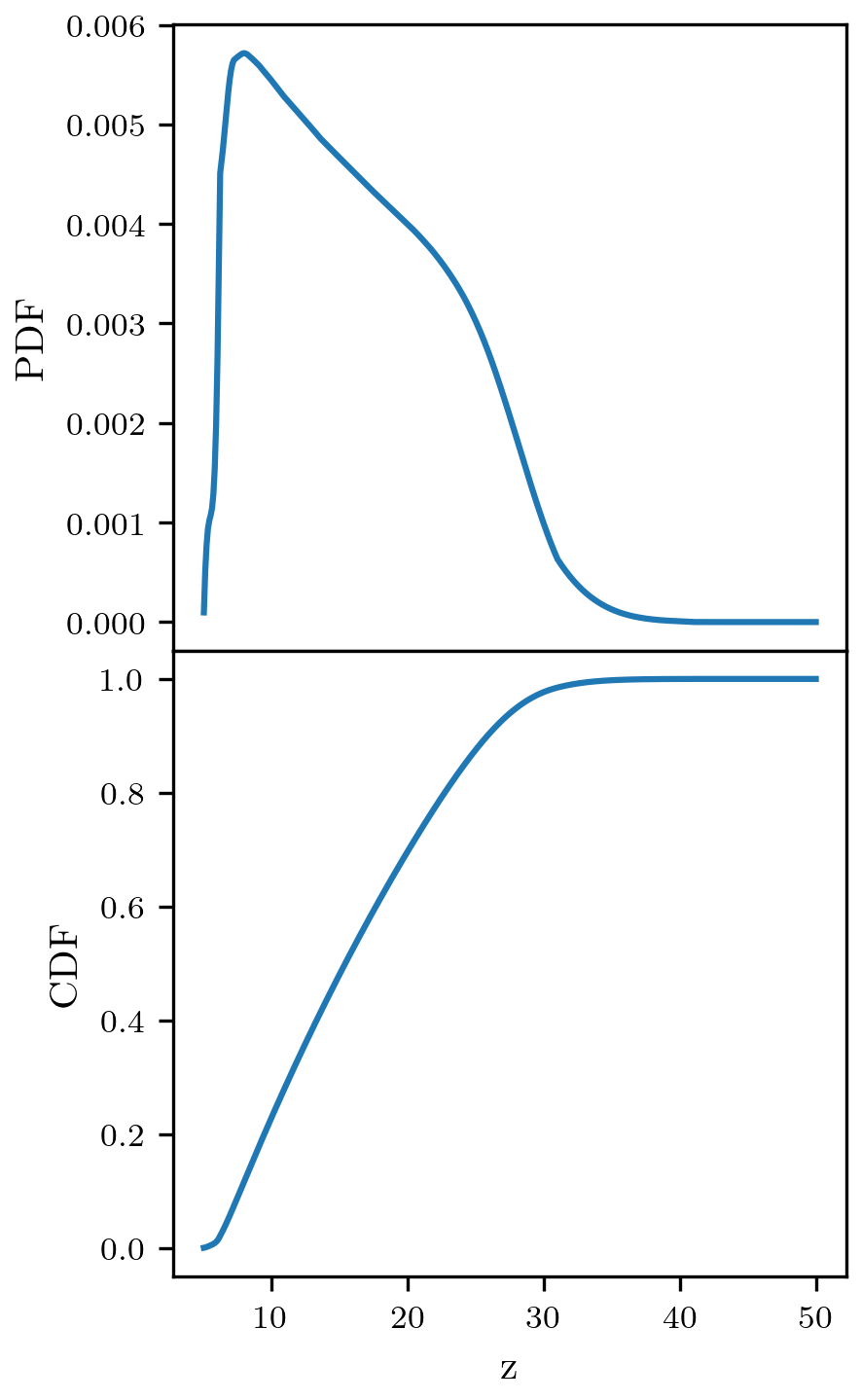}
    \caption{\textbf{Top panel:} The probability distribution calculated from the difference between the maximum and minimum signal temperatures in the \cmGEM~training data set using \cref{eq:resamp-dif} and \cref{eq:resamp-prob}. \textbf{Bottom panel:} The cumulative distribution function~(CDF) corresponding to the probability distribution in the top panel. We use this CDF to resample the training Global 21-cm signals in order to capture the variation at low redshifts across the distribution and allow the emulator to better learn this behaviour.}
    \label{fig:PDF_CDF}
\end{figure}

Where the variation in the signal at a given redshift across the training data set is large the probability distribution is also large (see \cref{fig:PDF_CDF}). We then calculate the corresponding cumulative distribution function~(CDF) and use inverse transform sampling to produce a new redshift distribution with a high sampling rate in regions of high variation. For each signal we can then perform an interpolation to get the corresponding $\delta T$ values. 

\subsection{Output and Input Normalisation}

Neural networks typically perform better when the outputs and inputs are of order unity and uniformly distributed. Hence it is typical to manipulate the data sets via logarithms, normalisation and/or standardisation to improve performance.

\begin{figure}
    \centering
    \includegraphics{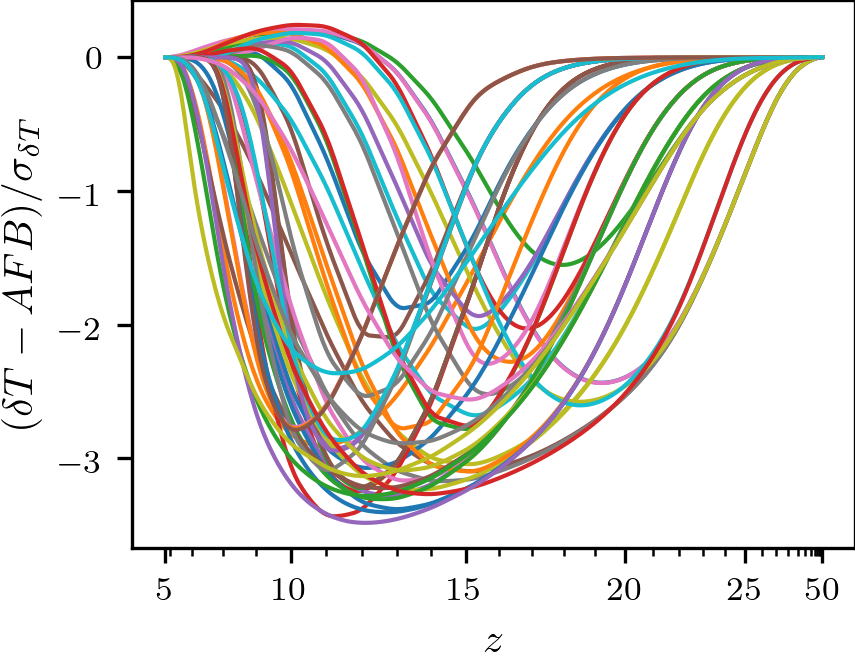}
    \caption{The equivalent signals from \cref{fig:example_signals} after pre-processing. Subtraction of the AFB and the subsequent resampling mean that the important information encoding the dependence on the astrophysical parameters is retained and appropriately emphasised in the training data. Here we have plotted the resampled redshift data points as being uniformly distributed since this is how the network is set up to interpret the input. The following division by the standard deviation across the training data set scales the signals to order unity without changing the physically significant value of $\delta T(z)=0$ where the spin temperature of the neutral hydrogen is equivalent to the radio background temperature. Minor ticks are at intervals of one on the x-axis.}
    \label{fig:ppSignals}
\end{figure}

After subtracting the AFB and resampling our signals we also proceed to divide the signals by the standard deviation across the training data set. This type of scaling was motivated by the typically used standardisation technique however we wanted to ensure that when scaling our signals a value of $\delta T = 0$ remained as 0 because it holds physical meaning~(an equivalence between the spin temperature and radio background, $T_s = T_r$). The signals shown in \cref{fig:example_signals}, as seen by the neural network, after pre-processing are shown in \cref{fig:ppSignals}. 

For the input redshift distribution we transform our resampled redshifts back onto a uniform distribution between 0 and 1, before they are input into the network, using the CDF detailed in the previous section. It is the combination of resampling and uniform redshift input that ensures the neural network `sees' `stretched' signals as in \cref{fig:ppSignals}. This technique allows the neural network to interpolate the signal at redshifts it hasn't been trained on to a higher degree of accuracy where the signals vary greatly than if we had used uniform sampling.

For the other input astrophysical parameters we use a Min-Max normalisation scaling each feature between 0 and 1. For example, considering the distribution of the CMB optical depth, $\boldsymbol{\tau}$ in our training data as a vector we normalise it such that
\begin{equation}
    \boldsymbol{\tilde{\tau}} = \frac{\boldsymbol{\tau} - \boldsymbol{\tau}_\mathrm{min}}{\boldsymbol{\tau}_\mathrm{max} - \boldsymbol{\tau}_\mathrm{min}}.
\end{equation}
The decision to use this type of normalisation was arrived at after testing standardisation, Min-Max normalisation and division by the max values for the input parameters while maintaining the physically motivated pre-processing for the signal temperatures detailed in the above sub-sections. 

For $f_X$, $f_*$ and $V_c$ the distributions are uniform in log-space and so we perform the Min-Max normalisation on the logarithm of these variables and use these as our inputs.

\subsection{$x_{HI}$ Pre-processing}
\label{neutral frac pp}

As discussed we provide provision in \name~to emulate the neutral fraction of hydrogen as a function of redshift. For this network the pre-processing just involves resampling of the signals since:
\begin{itemize}
    \item The equivalent AFB for the neutral fraction has a value of 1 at all redshifts and to subtract this from our training data set would invert our signals providing no benefit to training.
    \item The neutral fraction has a value between 0 and 1 by definition and so we do not need to normalise the output of the network to be of order unity.
\end{itemize}

\begin{figure}
    \centering
    \includegraphics{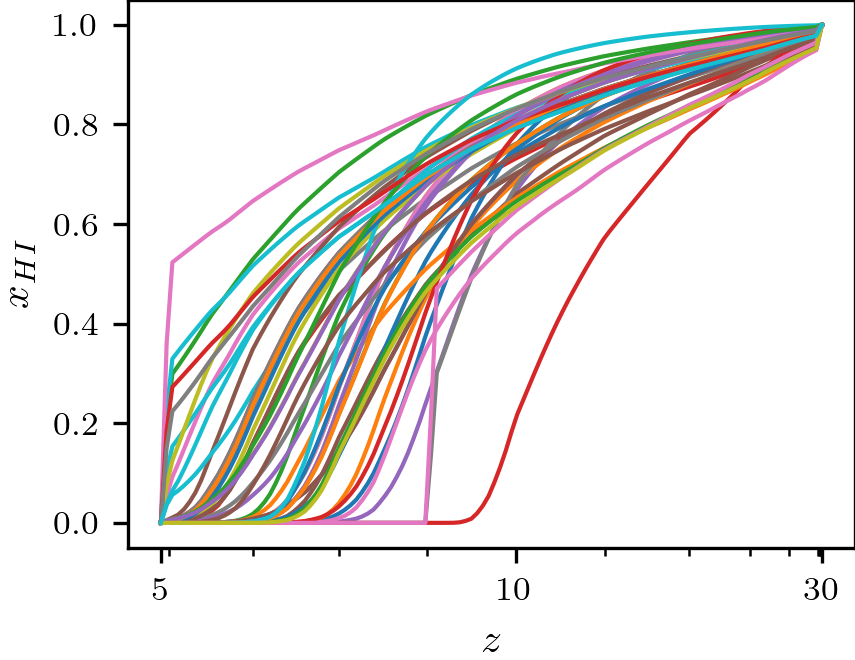}
    \caption{The equivalent neutral fraction histories from \cref{fig:xhi_example_signals} after pre-processing. For the neutral fraction histories, since the signals are already of order unity and subtraction of the equivalent AFB would not be beneficial, the pre-processing just involves resampling of the signals around regions of high variation. As with \cref{fig:ppSignals} the minor ticks are at intervals of one on the x-axis.}
    \label{fig:xHIppSignals}
\end{figure}

The benefits of performing resampling for the neutral fraction histories are the same as for the Global signal network. It allows the network to learn the variation in the training models and interpolation across redshift with a higher degree of accuracy. We perform the resampling with the equivalent of \cref{eq:resamp-dif} and \cref{eq:resamp-prob}. \Cref{fig:xHIppSignals} shows the same set of neutral fraction histories as in \cref{fig:xhi_example_signals} after pre-processing.

\section{Neural Network Structure}
\label{structure}

As stated, the goal with \name~is to maintain a simple network that is highly accurate without having to use dropout, regularisation, batch normalisation etc. However, in the design of any neural network the optimizer, the architecture, loss function, activation function and learning rate are core considerations. 

\subsection{Architecture}

Dropout \citep{Srivastava2014} and the commonly used L1 and L2 regularisation are typically employed to prevent overfitting where the network learns the training data to such a high degree of accuracy that it is unable to generalise. Overfitting is generally a result of using a neural network that is too big and has an excessive number of layers and nodes. On the other hand a network that is too small often produces poor quality predictions and consequently the aim is to produce a `reasonably' sized network. The scope of what constitutes a reasonably sized network is dependent on the number of input/output nodes, the variation in the training data and the complexity of the relationship between the inputs and outputs.

By using the novel approach of having redshift as an input to the network both our Global signal and neutral fraction emulators have, in the case of the \cmGEM~data, eight input nodes and one output node meaning that our network can remain small in size. Additionally, we have made a significant effort to simplify the problem with physically motivated pre-processing which also helps to reduce what constitutes as `reasonable' size for our networks.

\name~is set up in such a way that the number of layers and layer sizes can be adjusted by the user. As a result we do not provide a prescription of what constitutes a `reasonable' size as this may not be pragmatic. We note, however, that a significant effort can be undertaken to determine the optimum `reasonable' architecture that maximises accuracy and that this can also be impractical. Instead we suggest that as a minimum requirement a `reasonable' architecture for a trained \name~model should meet the following criteria:
\begin{itemize}
    \item The network should not over fit the training data otherwise the predictive power will be lost.
    \item The network should have an average accuracy $\lesssim 10$ per cent the noise of a typical Global 21-cm experiment (see \cref{accuracy} for a further discussion).
\end{itemize}

Based on the above criteria, the size of our input and output layers and a minimal exploration starting from a small network, trained with the pre-processed signals, and increasing the size until our accuracy criteria was met without overfitting (see \cref{app:overfitting}) we use a network with 3 hidden layers all of size 16 for both the Global signal and neutral fraction emulation in this paper.

\begin{figure}
    \centering
    \includegraphics{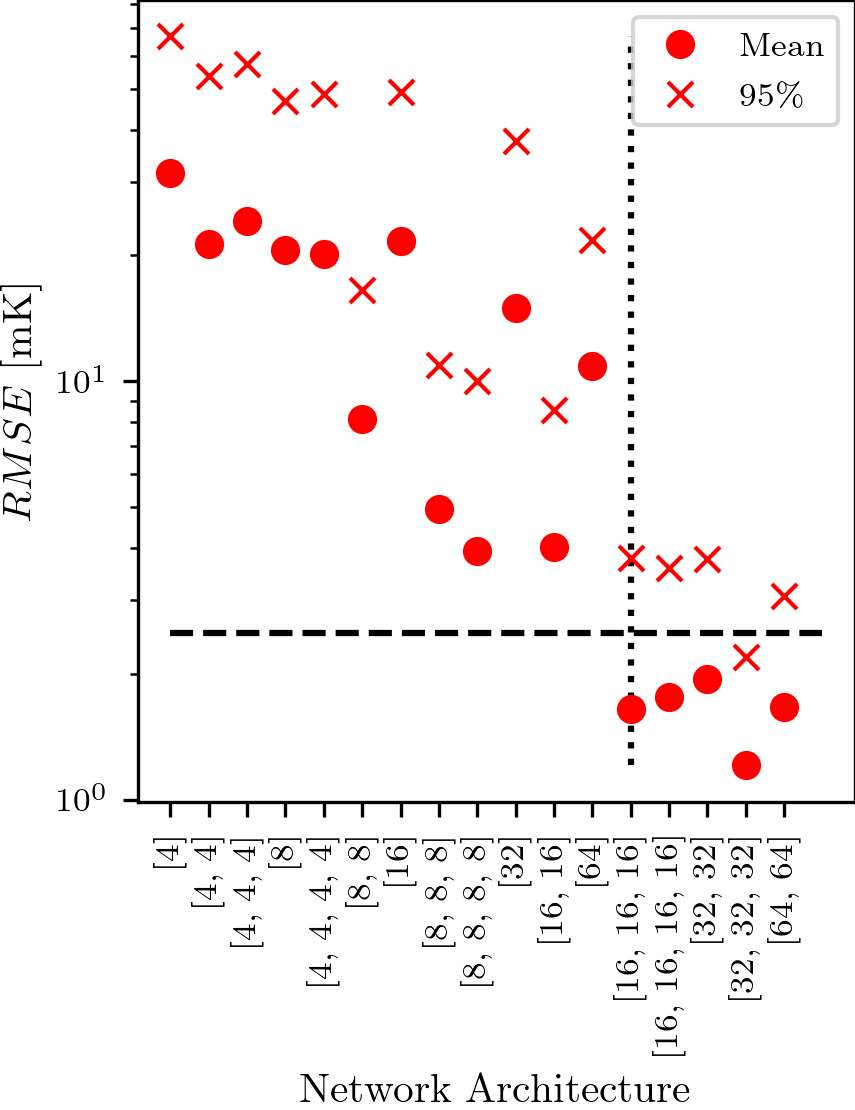}
    \caption{The mean and 95 percentile $RMSE$ (see \cref{accuracy}) for a set of different network architectures trained for 12h (approximately 250 epochs) on a HPC with the \cmGEM~Global signal training data and assessed with the corresponding test data. The architectures have between 1 and 4 layers of varying sizes between 4 and 64 nodes. They are ordered based on the number of weights in the network (equivalent to the number of connections) as this is a useful measure of network size and an indication of predictive power. The graph is used to determine a `reasonable' architecture considering the practical target accuracy of on average 10 per cent the expected noise of a Global 21-cm experiment (here illustrated by the black dashed line at 2.5~mK). Throughout the rest of the paper we use a network with 3 layers each consisting of 16 nodes which is the first to produce a mean value within our target accuracy. Our choice is highlighted with a dotted vertical line.}
    \label{fig:architecture}
\end{figure}

\Cref{fig:architecture} illustrates the processes used to determine our architecture for the Global network. We consider a set of different network sizes with one to four layers and 4, 8, 16, 32 or 64 nodes in each layer. For each of the tested networks we run a `full', 12 hours on a HPC equating to approximately 250 epochs using the full \cmGEM~training data, training of \name. We then assess the accuracy of the trained models using the $\approx1,700$ testing models in the \cmGEM~dataset. We compare the mean and 95 percentile $RMSE$ (see \cref{accuracy}) for each architecture. We find that a network of size [16, 16, 16] is the first to meet our target accuracy of on average 10 per cent the expected noise in a Global 21-cm experiment. 

While we may be able to achieve a better accuracy with a larger network this pragmatic approach leads to a sufficiently accurate network for physical signal modelling in the data analysis pipeline of a Global experiment like REACH. We also note that a smaller network can be evaluated faster than a larger architecture and that this is important when we are making multiple evaluations inside a nested sampling loop.

\subsection{Loss Function and Learning rate}

\name~uses the Mean Squared Error, typical for a regression network, as the loss function. In the case of the Global signal network is given by
\begin{equation}
    MSE = \frac{1}{N} \sum_{i=0}^N (\delta T_\mathrm{sim, i}(z_i) - \delta T_\mathrm{pred, i}(z_i))^2
\end{equation}
where $N$ is a batch size equivalent to the number of redshift data points in each signal. $\delta T_\mathrm{sim}(z)$ is the simulated signal temperature at a given redshift and $\delta T_\mathrm{pred}(z)$ is the emulated equivalent. \name~trains the neural networks in batches primarily to prevent memory related issues since the training data can be large~($\approx 27,000$ models times $451$ redshift points $\approx 12$ million data points for the \cmGEM~data). We find that a reasonable batch size is equal to the number of redshift data points in each model.

For the \cmGEM~data and the \name~framework we determine an effective learning rate to be 0.001. As with the architecture the learning rate can be adjusted by the user of \name~to meet the requirements of the data that they are training on.

\subsection{Optimizer}

The neural network optimizer is used to change the networks hyper-parameters to minimise the loss function. There is a number of different optimizers available~\citep{Ruder2016} and the choice can be dependent on the complexity of the problem and loss surface. A more robust optimizer is less likely to fall into and get stuck in local minima when training the network resulting in more accurate emulation. Therefore the choice of optimizer is important in designing an effective emulator. However, since \name~is designed to minimise the complexity of the relationship between the inputs and outputs and a MSE loss surface is relatively smooth\footnote{This can be assessed with a plot of the loss vs epoch number during training. We find that for the results presented in \cref{results} the surface is smooth up until the loss has plateaued and training is complete at which point we see noise like behaviour.} our choice is less consequential. We use, therefore, the commonly applied ADAM~\citep{Kingma2014} optimizer which is a momentum based modified stochastic gradient descent algorithm. 

\subsection{Activation Functions}

For both the Global signal and neutral fraction history network, we use a \textit{tanh} activation function in the hidden layers which can range between (-1, 1). However, for our final layer in the Global signal network we use a linear activation since the pre-processed temperature can be positive or negative and range between approximately $-4$ and $0.5$. Similar consideration is given to the output layer in the neutral fraction network where we use a \textit{ReLU}~(Rectified Linear Unit) activation which ensures that the output is always positive. Again the activation functions can be changed by the user of \name~to meet the requirements of their data. We note that the above output layer activations are designed to prevent unphysical outputs and that this is a crucial consideration for any user.

\section{Results}
\label{results}

\subsection{Emulation Time}
\label{timing}

In \cite{Cohen2020} the reported average time taken per signal with \cmGEM~is 160~ms when emulating a set of signals in a vectorised call. Here we compare the speed of \cmGEM~and \name~by emulating the 1,703 test models and taking an average time per signal. The tests are performed with \textsc{matlab} and \textsc{python} respectively on the same computer with the following processors: Intel® Core™ i3-10110U CPU @ 2.10GHz × 4. For \cmGEM~we make a vectorised call to the emulator as this results in a quicker performance then repeated single calls. We use a for loop to repeatedly call \name~which currently doesn't support such vectorised calls as they are not needed for physical signal modelling in a nested sampling loop.

For \name~we record a total time of 2.29 s and a corresponding average time per signal of $1.3 \pm 0.01$ mK. In comparison when emulating the same signals in a vectorised call with \cmGEM~we record a total time of 227.18 s and an average time per signal of 133~ms. We therefore achieve a factor of 102 improvement in emulation time with \name. We note that when using the \textsc{pymatbridge} (\url{https://github.com/arokem/python-matlab-bridge}) \textsc{python} wrapper for \textsc{matlab} the average time take to run a single prediction with \cmGEM~using a vectorised call is comparable to a direct call in \textsc{matlab}.

\subsection{Measuring Accuracy}
\label{accuracy}
In this section we primarily consider the accuracy of the Global signal emulator because the neutral fraction network has a similar design. We note, as previously stated, that the relationship between the neutral fraction and the astrophysical parameters is expected to be simpler and therefore easier to learn.

To assess the accuracy of \name~when emulating a Global 21-cm signal simulation we use a combination of two metrics; the root mean squared error~($RMSE$) and the normalised $RMSE$ given in \cite{Cohen2020} as
\begin{equation}
    \widetilde{RMSE} = \frac{RMSE}{max|\delta T_\mathrm{sim}(z)|},
    \label{eq:gemloss}
\end{equation}
where
\begin{equation}
    RMSE = \sqrt{\frac{1}{N} \sum_{i=0}^N(\delta T_\mathrm{sim, i}(z_i) - \delta T_\mathrm{pred,i}(z_i))^2}.
\end{equation}
For the neutral fraction network $\widetilde{RMSE}$ and ${RMSE}$~(with $x^{HI}_\mathrm{sim}(z)$ and the equivalent for the emulation in place of temperature) are equal since $max|x^{HI}_\mathrm{sim}(z)| = 1$. We assess the accuracy in the uniform redshift space and consequently our assessment is independent of the loss function used for training.

$\widetilde{RMSE}$ is a dimensionless quantity used by \cmGEM~and by assessing the quality of our network with this metric we can make direct comparisons between the two emulators. We also want our emulator to have an accuracy significantly lower than the expected noise floor of Global 21-cm experiments. This is required if the emulator is to be used to confidently model the Global 21-cm signal and draw conclusions about the astrophysics during the CD and EoR. To assess this accuracy requirement we can use the dimensionful $RMSE$ metric. 

As highlighted in the previous section, we suggest an average accuracy of $\lesssim 10$ per cent the expected noise of a Global 21-cm experiment such as REACH, equivalent to an $RMSE\lesssim 2.5$~mK, to be a sufficient limit. Since the accuracy of emulation is a function of the bandwidth, we report the accuracy across the entire range of the simulations $z = 5-50$~($z=5-30$ for the neutral fraction network) and across the expected REACH Phase I bandwidth of $z = 7 - 28$ (REACH Collaboration 2021 (in prep.)). The target is demonstratively achievable (see following section). It is also a practical target, if we want to use \name~for physical signal modelling, given that the noise in a 21-cm experiment has a fundamental effect on our confidence in any astrophysical parameter values inferred from the data and that this will likely be larger than the uncertainty introduced from \name.

\subsection{Global 21-cm Signal}

\begin{figure}
    \centering
    \includegraphics{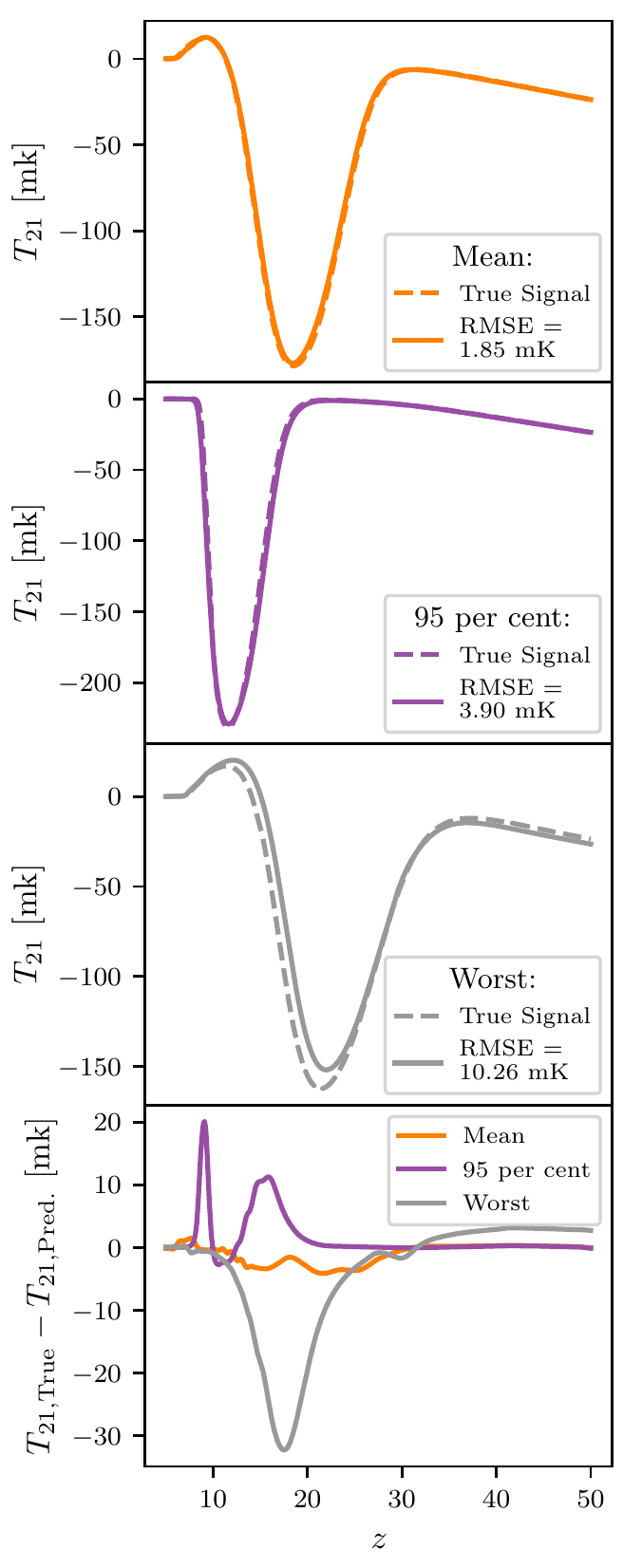}
    \caption{The top three panels show the mean, 95 percentile and the worst emulations respectively, based on the $RMSE$, for the Global 21-cm signal across the entire test set of 1,703 models. The bottom panel shows the difference between the simulations and predictions as a function of redshift. Full details of the accuracy of the emulation can be found in \cref{tab:full_results} and a discussion can be found in the text.}
    \label{fig:bestWorstT}
\end{figure}

\cref{fig:bestWorstT} shows that the mean $RMSE$ value across the redshift range $z=5 - 50$ is 1.85 mK and that the maximum value is 10.26 mK. Further, \cref{tab:full_results} shows that performing the same calculation of the $RMSE$ inside the REACH band, $z=7-28$ gives a mean value of $2.52$ mK which is very close to the desired $2.5$~mK limit. We also report in the table the $RMSE$ value for which $95$ per cent of the models have a value smaller than or equal to. In the REACH band this equates to $5.37$ mK. For all of the reported results the 95 percentile is significantly lower than the maximum values (a factor of 3 for the Global signal and a factor of approximately 1.5 to 2 depending on bandwidth for the neutral fraction histories). This means that out of a set of $1,703$ Global signals only 85 have $RMSE$ values above $3.90$ mK across the band $z = 5-50$ for example. We note that the values reported, averages across redshift ranges, in the REACH band are generally higher than across the whole redshift range because the REACH band excludes redshifts $\gtrsim 30$ where the emulation is expected to be very precise.

\Cref{paramVsError} shows the explored parameter space for the Global 21-cm signal in the \cmGEM~test data set and the corresponding error when emulating the signals with \name.

Finally, in \cref{tab:full_results} we also report the $\widetilde{RMSE}$ values in both the REACH band and across the whole redshift range. \cite{Cohen2020} report similar results for \cmGEM~and particularly we note that, when training and testing on the same data sets, we recover a mean $\widetilde{RMSE}$ of 1.12 per cent compared to 1.59 per cent when using \cmGEM. Similarly we report a maximum value of $6.32$ per cent in comparison to the value of $10.55$ per cent reported by \citeauthor{Cohen2020} for \cmGEM. This further demonstrates that \name~can achieve a high degree of accuracy in its emulation.

\subsection{Neutral Fraction}

\begin{figure}
    \centering
    \includegraphics{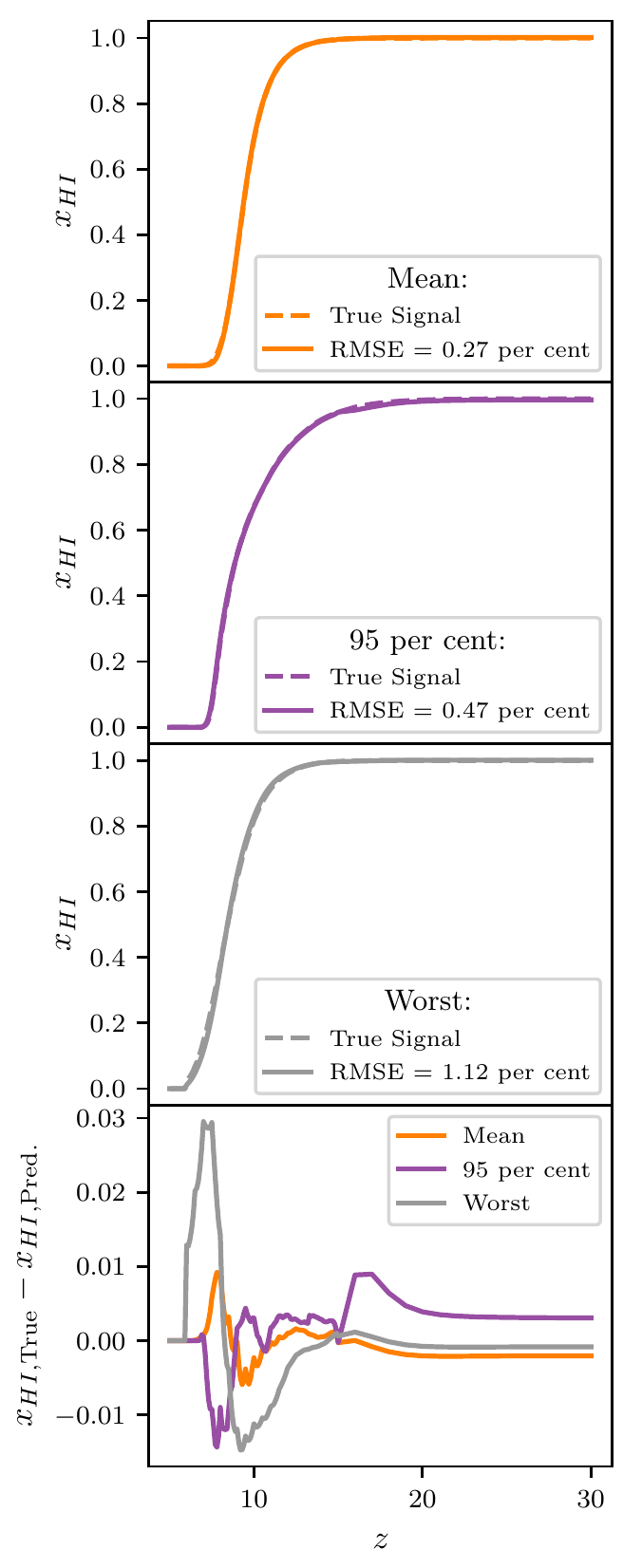}
    \caption{The top three panels show the mean, 95 percentile and the worst emulations respectively for the neutral history across the entire test set of 791 models and the difference between the simulations and predictions is shown in the bottom panel. The level of accuracy here is higher than that for the Global signal, despite using a similar pre-processing and identical network, demonstrating that the relationship between the astrophysical parameters, redshift and the network output is simpler here.}
    \label{fig:bestWorstxHI}
\end{figure}

For the neutral fraction history network we show similar results. \Cref{fig:bestWorstxHI} demonstrates the quality of the emulation with the mean, 95 percentile and the worst results when emulating the neutral fraction and these values are detailed in \cref{tab:full_results}. The results generally are of higher quality than that for the Global signal and, noting that the pre-processing for the two networks is near identical and the networks themselves are of the same size, this supports the understanding that the relationship between the inputs and outputs is simpler here. In the band $z = 5 -50$ only 39 of 791 test models have $\widetilde{RMSE} \geq 0.47$ per cent.

\begin{table*}
    \centering
    \begin{tabular}{c|c|c|c|c|c}
         & & \multicolumn{2}{|c|}{Global Signal} & \multicolumn{2}{|c|}{Neutral Fraction} \\
         \hline
         & & $z = 5- 50$& $z = 7- 28$& $z = 5- 30$& $z = 7- 28$ \\
         \hline
         \multirow{4}{4em}{$RMSE$}& Minimum &  0.30 mK &  0.31 mK& 0.09 per cent & 0.08 per cent \\
         & Mean & 1.85 mK& 2.52 mK& 0.29 per cent & 0.26 per cent \\
         & $95^{th}$ percentile& 3.90 mK& 5.37 mK& 0.47 per cent& 0.44 per cent\\
         & Maximum & 10.26 mK& 15.10 mK& 1.12 per cent & 0.65 per cent \\
         \hline
         \multirow{4}{4em}{$\widetilde{RMSE}$}& Minimum & 0.21 per cent & 0.26 per cent& -- & -- \\
         & Mean & 1.12 per cent & 1.53 per cent & -- & -- \\
         & $95^{th}$ percentile& 2.41 per cent & 3.22 per cent & -- & -- \\
         & Maximum & 6.32 per cent & 9.31 per cent & -- & -- \\
    \end{tabular}
    \caption{Detailed results of the emulation using \name~and the \cmGEM~training and test data for both the Global signal and the neutral fraction history. We find that \name~ achieves the desired accuracy of on average $\approx 10$ per cent the expected noise of a typical Global 21-cm experiment (equating to $\approx 2.5$~mK in the REACH band of $z=7-28$). Of note are the recorded $95$ per cent percentiles, the $RMSE$ for which $95$ per cent of the models have values smaller than or equal to, which are significantly lower than the maximum $RMSE$ values. A discussion comparing the results of \cmGEM~and \name, in terms of $\widetilde{RMSE}$, can be found in the text. Briefly we find that our Global signal emulator has a maximum $\widetilde{RMSE}$ approximately half that achieved with \cmGEM. For the neutral fraction $RMSE = \widetilde{RMSE}$ and so we only report one set of results. We find a higher degree of accuracy here with an identical network and similar pre-processing indicating a simpler relationship.}
    \label{tab:full_results}
\end{table*}

\section{Conclusions}
\label{conclusions}

\name~uses a novel approach to emulate, with neural networks, the Global 21-cm signal and the evolution of the neutral fraction during the CD and EoR by considering redshift as an input to the neural networks alongside the astrophysical parameters. In tandem with this reparameterisation of the problem we use a predominantly physically motivated pre-processing for both the Global signal and neutral fraction. We subtract from the Global signals an astrophysics free baseline which obviates the need for the network learning a non-trivial but well-understood relationship at high redshift. We then resample both the Global signals and neutral fractions so that the regions which vary significantly across the training data sets can be better characterised by the networks.

The above framework allows the complex relationships between the astrophysical parameters and the Global signal or neutral fraction history as functions of redshift to be effectively learnt with small neural networks. Each Global signal of 451 redshift data points can be emulated in on average $1.3$~ms. We note that this is a factor of approximately $102$ improvement on the 133~ms we record with \textsc{matlab} when predicting the same signals on the same computer with \cmGEM.

We demonstrate the effectiveness of \name~by using the \cmGEM~training and testing data. This allows for a direct comparison between our results and the results of \cmGEM. We find that \name~can emulate to a higher degree of accuracy the Global 21-cm signal than \cmGEM~with a maximum normalised $RMSE$ of 6.32 per cent in comparison to 10.55 per cent over the range $z = 5 -50$. We also demonstrate that \name~can emulate a Global 21-cm signal to, on average, less than 10 per cent the expected noise of a Global 21-cm experiment like REACH.

Finally, \name~is a flexible \textsc{python} package that can be easily retrained on updated models with new astrophysical dependencies. For example additional astrophysical phenomena such as Lyman-$\alpha$ heating \citep{Reis2021} or additional radio background produced by galaxies (or an indeterminate synchrotron-like source) \citep{Fialkov2019, Reis2020} can be incorporated and easily trained upon. While the results achieved with the \cmGEM~data are impressive, the novelty of \name~is in its flexibility, incorporation of redshift as an input and physically motivated pre-processing. Particularly the final two points allow for an accurate mapping from parameters to temperature with a single neural network reducing the points of failure and need for excessive fine-tuning.

\section*{Acknowledgements}

HB acknowledges the support of the Science and Technology Facilities Council (STFC) through grant number ST/T505997/1. WH and AF were supported by Royal Society University Research Fellowships. EA was supported by the STFC through the Square Kilometer Array grant G100521.

\section*{Data Availability}

The Global 21-cm signals used in this paper are publicly available at \url{https://doi.org/10.5281/zenodo.4541500} and provided by \cite{Cohen2021Data}.



\bibliographystyle{mnras}
\bibliography{References.bib} 




\appendix

\section{Calculating the Astrophysics Free Baseline}
\label{appendixA}

To approximate the astrophysics free baseline~(AFB) we need to consider the physics defining the signal structure during the period dominated by collision coupling. During this period neutral hydrogen atoms collide with other neutral hydrogen atoms, protons and electrons. The spin temperature of hydrogen is coupled to the gas temperature, $T_K$, via the collisions and that temperature cools adiabatically at a faster rate than the background radiation, $T_r$. The spin temperature, $T_s$, which encodes the number of hydrogen atoms in the two hyperfine levels of the ground state \citep{Furlanetto2006} during this period is given by
\begin{equation}
    \frac{1}{T_s} = \frac{1/T_r + x_c/T_K}{1+x_c}
\end{equation}
where $x_c$ is the collisional coupling coefficient. For our approximation of the AFB calculated here we use a reference value for the gas temperature of $T_{K\mathrm{, ref}} = 33.7340$~K at $z_\mathrm{ref} = 40$ from the simulations used to produce the training and test data sets. We then scale $T_{K\mathrm{, ref}}$ adiabatically using
\begin{equation}
    T_K = T_{K\mathrm{, ref}} \frac{(1 + z)^2}{(1+z_\mathrm{ref})^2}
\end{equation}
to get $T_K$ as a function of redshift.

The coupling is dominated by H-H collisions and so we only consider these in our simulation. The coupling coefficient for this interaction is given by \cite{Furlanetto2006}
\begin{equation}
    x_c^{HH} = \frac{n_H \kappa_{10}^{HH} T_*}{A_{10} T_r},
\end{equation}
where $\kappa_{10}^{HH}$ is the rate coefficient for the spin deactivation of neutral hydrogen, $T_*$ is the energy defect and $A_{10}$ is the spontaneous emission coefficient of the 21-cm transition. $n_{H}$ is the relative number density of neutral hydrogen given by as
\begin{equation}
    n_H = 3.40368\times 10^{68}~\frac{\rho_c}{m_p} (1 - Y) \Omega_b (1 + z)^3
\end{equation}
where $Y=0.274$ and is the Helium abundance by mass, $\rho_c$ is critical mass density of the universe in $M_\mathrm{sol}/$cMpc$^3$, $m_p$ is the proton mass in $M_\mathrm{sol}$ and $\Omega_b$ the the baryon density parameter.

From the above we can then calculate $\delta T$ as
\begin{equation}
    \delta T = \frac{T_s - T_r}{1 + z}(1 - \exp(-\tau_{\nu_0})),
\end{equation}
where $\tau_{\nu_0}$ is the 21-cm optical depth of the diffuse IGM
\begin{equation}
    \tau_{\nu_0} = \frac{3 h c^3 A_{10} x_{HI} n_H}{32 \pi k_b T_s \nu_0^2 H(z)},
\end{equation}
where $\nu_0$ is the rest frequency of the 21-cm emission and $H(z)$ is the Hubble rate. In our calculation we use the same cosmological parameters that were used to generate the signals \citep[see][]{Cohen2020}. Here the neutral fraction, $x_{HI}$ has a value of 1 since there is no astrophysics involved in the AFB.

\section{Testing the neural networks for overfitting}
\label{app:overfitting}

\begin{figure}
    \centering
    \includegraphics{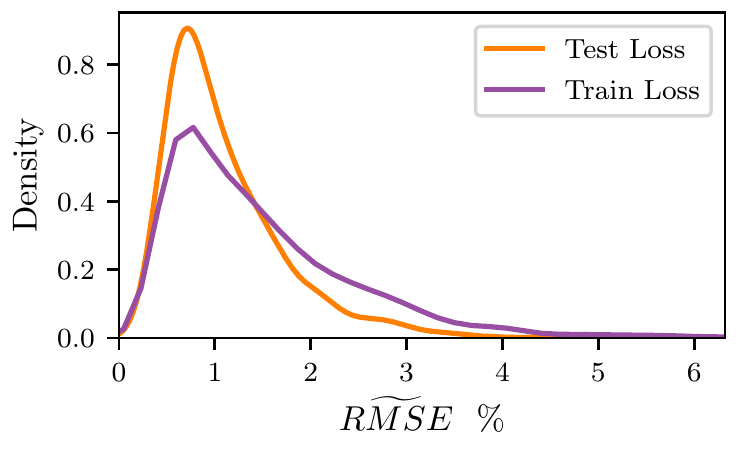}
    \caption{The probability density for the loss distribution found when emulating the training and test data sets for the Global 21-cm signal with \name.}
    \label{fig:globalsignaloverfitting}
\end{figure}

\begin{figure}
    \centering
    \includegraphics{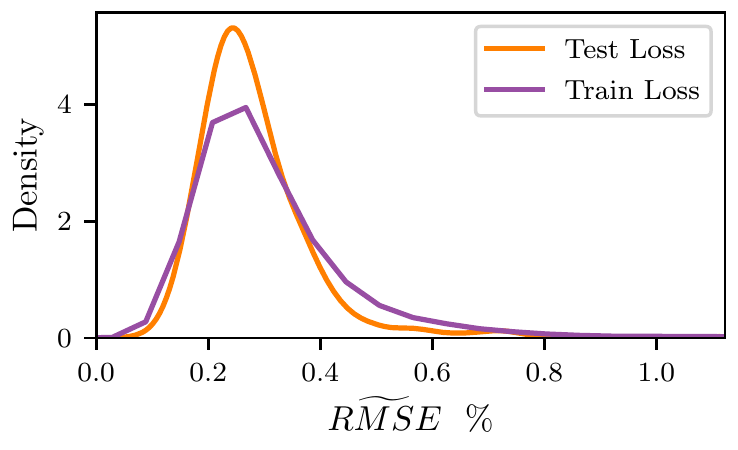}
    \caption{The probability density for the loss distribution found when emulating the training and test data sets for the neutral fraction with \name.}
    \label{fig:xhioverfitting}
\end{figure}

We can demonstrate that, for both the Global signal and neutral fraction, the chosen network sizes of 3 hidden layers of 16 nodes do not overfit the training data by comparing the distribution of loss values across the training and test data sets. This is shown in \cref{fig:globalsignaloverfitting} and \cref{fig:xhioverfitting} for the Global signal and neutral fraction respectively. Again, we have used a gaussian kernal density estimation to calculate continuous probability density curves from the discrete histograms of losses. We can see in both cases that the loss, evaluated with \cref{eq:gemloss}, for the testing and training data sets, when emulated with the trained neural networks, have similar distributions and can consequently conclude that the neural networks are not overfitting the training data. In the event that the training data was being overfit then the purple distribution, showing the training data losses, would peak to the left of the orange distribution, showing the test data losses, because the network would have learnt the training data to such a high degree of accuracy that it cannot generalise well to the testing data.

\section{Error Vs Parameter}
\label{paramVsError}

\begin{figure*}
    \centering
    \includegraphics{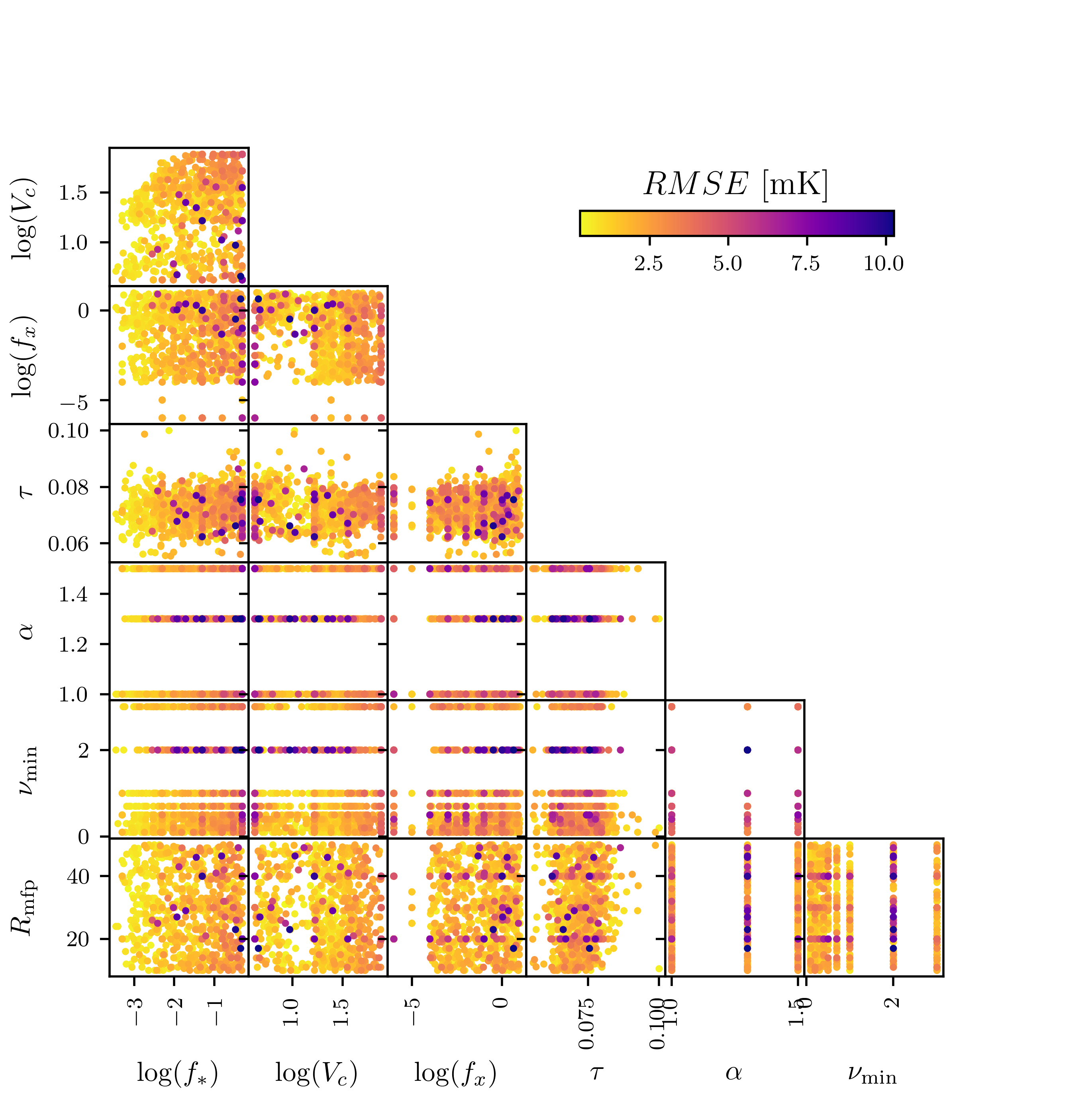}
    \caption{The parameter space explored by the \cmGEM~test data set. Each panel shows the $1,703$ models plotted as data points based on the corresponding astrophysical parameter values. They are coloured according to the $RMSE$ calculated when comparing the true signals to the emulation from \name~across the range $z = 5-50$.}
    \label{fig:paramVsError}
\end{figure*}

\Cref{fig:paramVsError} shows the parameter space explored in the \cmGEM~test data set as a scatter plot. The data points are coloured based on the $RMSE$ value calculated when comparing the corresponding true signal with the emulation, over the range $z = 5 - 50$, from \name. 

\begin{figure*}
    \centering
    \includegraphics{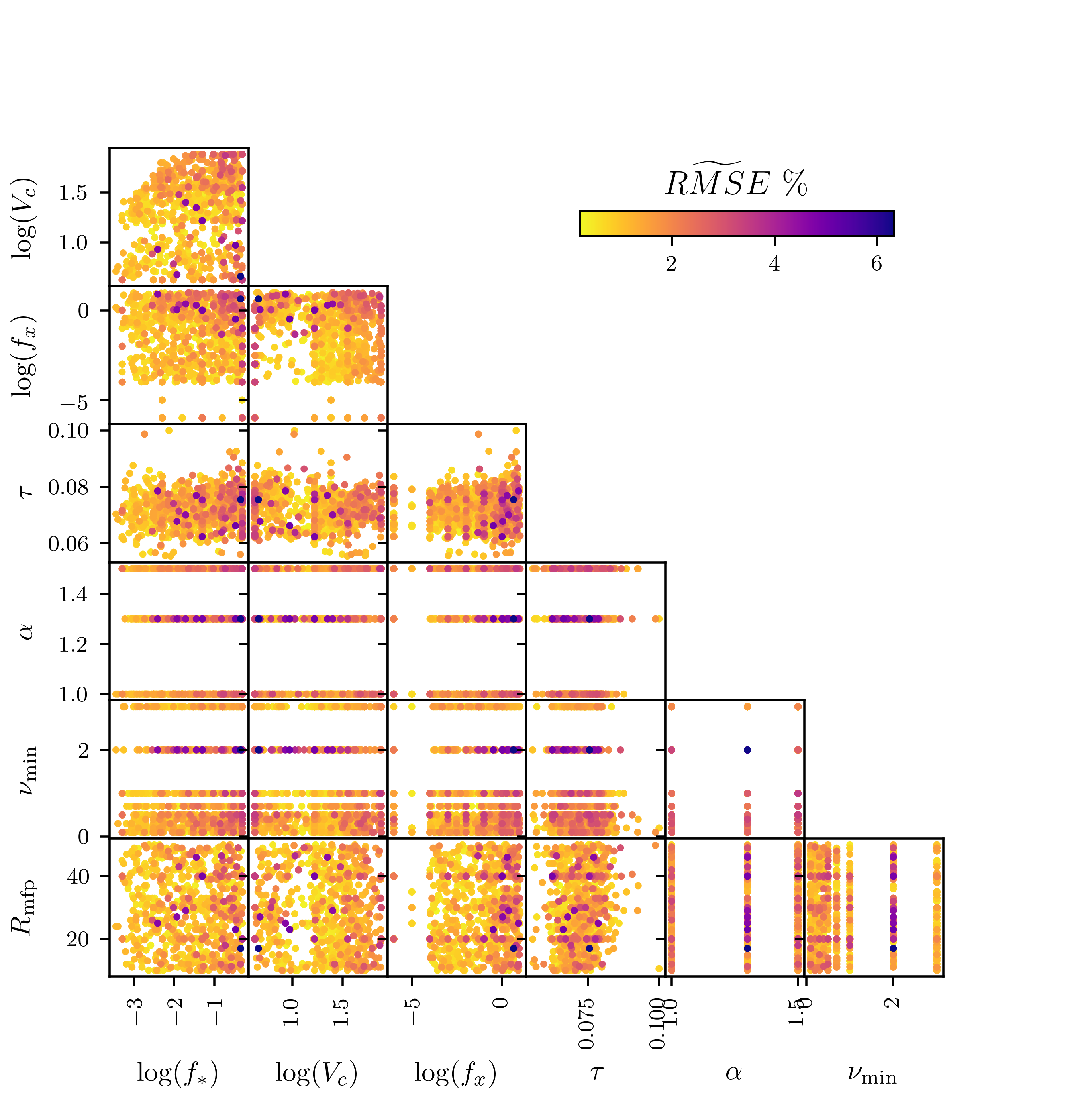}
    \caption{The equivalent of \cref{fig:paramVsError} with the data points colored based on the dimensionless $\widetilde{RMSE}$ error calculated across the band $z = 5 - 50$.}
    \label{fig:paramVsErrorGEM}
\end{figure*}

\Cref{fig:paramVsErrorGEM} shows the equivalent graph with the colours determined using the dimensionless $\widetilde{RMSE}$ metric across the band $z = 5 - 50$.

\bsp	
\label{lastpage}
\end{document}